\documentclass[twocolumn,tighten]{aastex631}
\usepackage{booktabs} 
\usepackage{array}    

\usepackage{amsmath}
\usepackage{xcolor}
\usepackage{wasysym}
\usepackage{multirow}
\usepackage{newtxtext,newtxmath}
\usepackage{afterpage} 
\usepackage[T1]{fontenc}
\usepackage{gensymb}

\usepackage{graphicx}	
\usepackage{amsmath}	
\usepackage{gensymb}    

\usepackage{verbatim}

\definecolor{my_color}{HTML}{3a18b1}
\definecolor{comments}{HTML}{800000}
\definecolor{black}{HTML}{000000}

\definecolor{new_color}{HTML}{CF0000}
\definecolor{new_black}{HTML}{000000}
\usepackage{xfrac}

\newcommand{\bedit}[1]{\textcolor{black}{#1}}

\newcommand{\be}{\begin{equation}}
\newcommand{\ee}{\end{equation}}

\submitjournal{AJ}

\shorttitle{Look what you made me accrete}
\shortauthors{Jenkins et al.}

\begin{document}

\title{Atmospheric Signatures of Common Envelope Evolution in White Dwarf Planets
}

\author[0000-0001-9827-1463]{Sydney A. Jenkins} 
\altaffiliation{NSF Graduate Research Fellow}
\affiliation{Department of Physics and Kavli Institute for Astrophysics and Space Research, Massachusetts Institute of Technology, 77 Massachusetts Avenue, Cambridge, MA 02139, USA}

\author[0000-0001-7246-5438]{Andrew Vanderburg}
\affiliation{Department of Physics and Kavli Institute for Astrophysics and Space Research, Massachusetts Institute of Technology, 77 Massachusetts Avenue, Cambridge, MA 02139, USA}

\author[0000-0001-5864-9599]{James Mang}
\altaffiliation{NSF Graduate Research Fellow}
\affiliation{Department of Astronomy, University of Texas at Austin, 2515 Speedway, Austin, TX 78712, USA}

\author[0000-0002-7733-4522]{Juliette Becker}
\affiliation{Department of Astronomy, University of Wisconsin-Madison, 475 N. Charter Street, Madison, WI 53706, USA}

\author[0000-0002-4404-0456]{Caroline V. Morley}
\affiliation{Department of Astronomy, University of Texas at Austin, 2515 Speedway, Austin, TX 78712, USA}

\author[0000-0001-7493-7419]{Melinda Soares-Furtado}
\affiliation{Department of Astronomy, University of Wisconsin-Madison, 475 N. Charter Street, Madison, WI 53706, USA}
\affiliation{Department of Physics, 2320 Chamberlin Hall, University of Wisconsin–Madison, 1150 University Avenue Madison, WI 53706-1390}

\author[0000-0003-0381-1039]{Ricardo Yarza}
\affiliation{Department of Astronomy and Astrophysics, University of California, Santa Cruz, CA 95064, USA}


\correspondingauthor{Sydney Jenkins}
\email{sydneyaj@mit.edu}



\begin{abstract}
The majority of confirmed exoplanets orbit within 1 au of a main-sequence (MS) star. When their stellar hosts evolve off the MS, many of these planets will be engulfed and destroyed, creating empty “forbidden” zones around the stars as they evolve to their final state as a white dwarf (WD). However, several confirmed and candidate WD planets have been found within \bedit{this forbidden zone.} 
Two formation scenarios have been proposed to explain the existence of these close-in planets: \bedit{high-eccentricity} migration and common envelope evolution (CEE). There are currently few observational tests to distinguish between these pathways. In this study, we investigate whether CEE could leave a detectable atmospheric signature. Using Modules for Experiments in Stellar Astrophysics ({\tt MESA}) models, we simulate an engulfed planet inspiraling into an AGB star, and allow the planet to accrete mass via Bondi-Hoyle-Lyttleton accretion. \bedit{Assuming a range of planet masses (1$-$13 M$_{\mathrm{Jup}}$) and accretion efficiencies (0.01$-$1.0), we find that the planet can accrete up to 48\% of its initial mass in the most extreme Eddington-limited scenario. Because this accreted material is enriched in hydrogen and helium, we expect it to decrease the planet's bulk metallicity. Using simulated emission spectra, we find that CEE can increase thermal emission by up to 9.0\% for a cool planet such as WD 1856 b. For lower accretion efficiencies (0.01$-$0.5), thermal emission increases between 0.1$-$3.6\%. This signature may be observable in the most favorable cases, providing} a potential new probe for investigating the dynamical history of close-in planets around WDs. 

\end{abstract}



\section{Introduction}
Though most known exoplanets orbit main-sequence stars, there is ample evidence that planets can survive their star’s transition off the main-sequence. For instance, more than a quarter of white dwarfs (WDs) have photospheric metal pollution \citep{koe2014} consistent with the accretion of planetary bodies \citep[e.g.,][]{jur2003, far2009, zuc2010}; many have dusty or gaseous discs \citep[e.g.,][]{zuc1987, gan2006, gan2007, far2009, roc2015, den2020, mel2020}; and several systems exhibit transiting debris clouds from disintegrating rocky bodies, such as WD 1145+017 \citep{van2015, gan2016}. Intact planets have also been found orbiting WDs \bedit{\citep{tho1993, luh2011, gan2019, van2020, bla2021, zha2024} and pulsars \citep[e.g.,][]{Wolszczan1992Natur, Spiewak2018MNRAS, Vleeschower2024MNRAS},} providing direct evidence that planets survive in post-main-sequence systems.



Theory suggests that the population of planets around WDs should be limited to wide-orbit planets. 
As a star evolves on the red giant branch (RGB) or asymptotic giant branch (AGB), it is expected to expand to $\sim$1$-$2 au, engulfing any close-in planets and carving out an empty "forbidden zone." 
Most of these engulfed worlds will be smaller terrestrial or sub-Neptune worlds, and will be quickly destroyed. However, based on planet demographics \citep[e.g.,][]{cum2008, fer2019, ful2021} and inspiral rates \citep{mus2012}, $\sim$10\% of solar-like stars are expected to engulf a 1$-$10 M$_{\mathrm{Jup}}$ planet as they evolve on the RGB or AGB \citep{oco2023}. Assuming all engulfed planets are destroyed, we expect the WD planet population to be dominated by wide-orbit ($\gtrsim$2 au) worlds, with only a small subset of these planets migrating inwards after the RGB/AGB phases are complete \citep{Debes2002, Mustill2014, Veras2020}. 

Recent observational evidence reveals that such close-in planets do exist: \citet{gan2019} discovered a WD with accretion consistent with an evaporating giant planet at 0.07 au and \citet{van2020} discovered a transiting planet, WD 1856+534 b (hereafter WD 1856 b), that orbits at just 0.0204 au. 
Two formation scenarios have been proposed that explain WD 1856 b's extreme orbit and that may be important for the wider WD planet population: high-eccentricity migration \citep[HEM, e.g.,][]{mun2020, oco2021} and common envelope evolution \citep[CEE,][]{cha2021, lag2021, mer2021}. In a HEM scenario, the planet starts at a wider orbit around the WD star and is driven to high eccentricities by either the Kozai-Lidov mechanism or planet-planet scattering. It then undergoes HEM and tidal circularization to reach its current orbit \citep{Veras2019, Veras2019b}. 

In a CEE scenario, the planet is engulfed during the host star’s transition off the main-sequence. The planet then inspirals into the star, injecting energy into the stellar envelope. If there is sufficient energy to expel the envelope before the planet fills its Roche lobe, then the planet may survive.  This likely requires some additional source of energy, such as a previous engulfed planet, recombination energy in the envelope, or convection during the core helium flash of the RGB phase \citep{lag2021, mer2021}. A population of candidate post-common envelope binaries containing brown dwarfs has been discovered \citep[e.g.,][]{max2006, beu2013, ste_2013, far_2016, par2017, cas2020, van2021}, suggesting that higher mass companions may survive engulfment. Additionally, a potential example of a surviving giant planet around a core helium-burning red giant was discovered by \citet{hon2023}. The planet, 8 Ursae Minoris b, orbits its host star at a separation of just 0.462 $\pm$ 0.006 au, within the predicted maximum radius of the star (0.7 au). Alternative explanations for 8 Ursae Minoris b's unique orbit are that a stellar merger modified its host star's evolution, that it is a second-generation planet, or that the host star is more massive than initially predicted \citep{che2024}.

Though both HEM and CEE provide a potential explanation for why we find close-in planets around WDs, there are few observational methods for distinguishing between the two formation pathways. In this work, we propose a new observational technique for constraining the dynamical history of close-in planets around WDs. In a CEE scenario, the planet will accrete a considerable amount of material from the star’s interior while it moves through the stellar envelope, potentially leaving a unique signature in the planet's atmosphere. 
We describe our CEE model in \S\ref{Model}, and explore potential complications to our model in \S\ref{sec:Complications}. In \S\ref{Results} we estimate how much mass an engulfed planet might accrete, and predict the resulting atmospheric signatures in the planet's emission spectrum. We then discuss conditions for planetary survival, directions for future work, and prospects for observational detection in \S\ref{Discussion} before concluding in \S\ref{Conclusion}. 

\section{Model Description}
\label{Model} 
To investigate how much mass a planet can accrete during CEE, we model an engulfed giant planet inspiralling into an evolving star. We describe our stellar and planetary models in \S\ref{Stellar_model} and \S\ref{Planet_model}, respectively. We then account for mass accretion from the stellar envelope, described in \S\ref{Bondi-Hoyle}. Our final set of simulations is described in \S\ref{Simulation}.

\subsection{Stellar Model}
\label{Stellar_model}
Giant planets are more abundant at wide orbital separations, with occurrence rates of $\approx$10$-$15\% between 1$-$5 au \citep{fer2019, ful2021} and $\approx$3\% within 1 au \citep{cum2008, may2011}. They are therefore more likely to be engulfed as their host star reaches its maximum expansion near the tip of the RGB or AGB (Gaibor et al. \textit{in prep}). 
In our simulations, we therefore assume that the host star has evolved to the tip of the AGB. 

\bedit{To} model the star's evolution, we use Modules for Experiments in Stellar Astrophysics \citep[{\tt MESA}, ][]{Paxton2011ApJS,Paxton2013ApJS,Paxton2015ApJS,Paxton2018ApJS,Paxton2019ApJS} version 24.08.1. We assume a non-rotating progenitor star with a mass of 2 M$_\odot$, consistent with WD 1856's progenitor mass, and use initial chemical abundances from \citet{asp2009} with Z$_{\rm base} $ = 0.0142. We model convection using the turbulent diffusive convection (TDC) formalism of  Mixing Length Theory \citep{kuh1986}, with a mixing length scale factor $\alpha_{\rm MLT} $ of 2.0. We also account for wind-driven mass loss by applying Reimers' mass loss prescription \citep{rei1975} from the zero-age MS through the RGB, and applying Blöcker's mass loss prescription \citep{blo1995} on the AGB, with scaling factors of 0.5 and 0.1, respectively. We use the OPAL opacity tables with the \texttt{a09} prefix \citep{asp2009}, consistent with our adopted solar abundances.

We evolve the stellar model from the zero-age main sequence through the RGB tip, the end of core helium burning, and into the thermally-pulsing AGB phase. \bedit{We then extract the stellar model at the tip of the first thermal pulse, to be used for our subsequent planet injection}. 
This 
captures the characteristic behavior of thermally-pulsing AGB evolution without the computational expense and numerical challenges of modeling later, more unstable pulses. 
Prior to the planet's inspiral, some additional energy would likely need to be added to the envelope in order to eject it before the planet is destroyed. However, we do not include this energy injection in our stellar model. See Gaibor et al. \textit{in prep} for a discussion of how this would impact planetary survival. 

\begin{figure}
\centering
\includegraphics[width=\columnwidth]{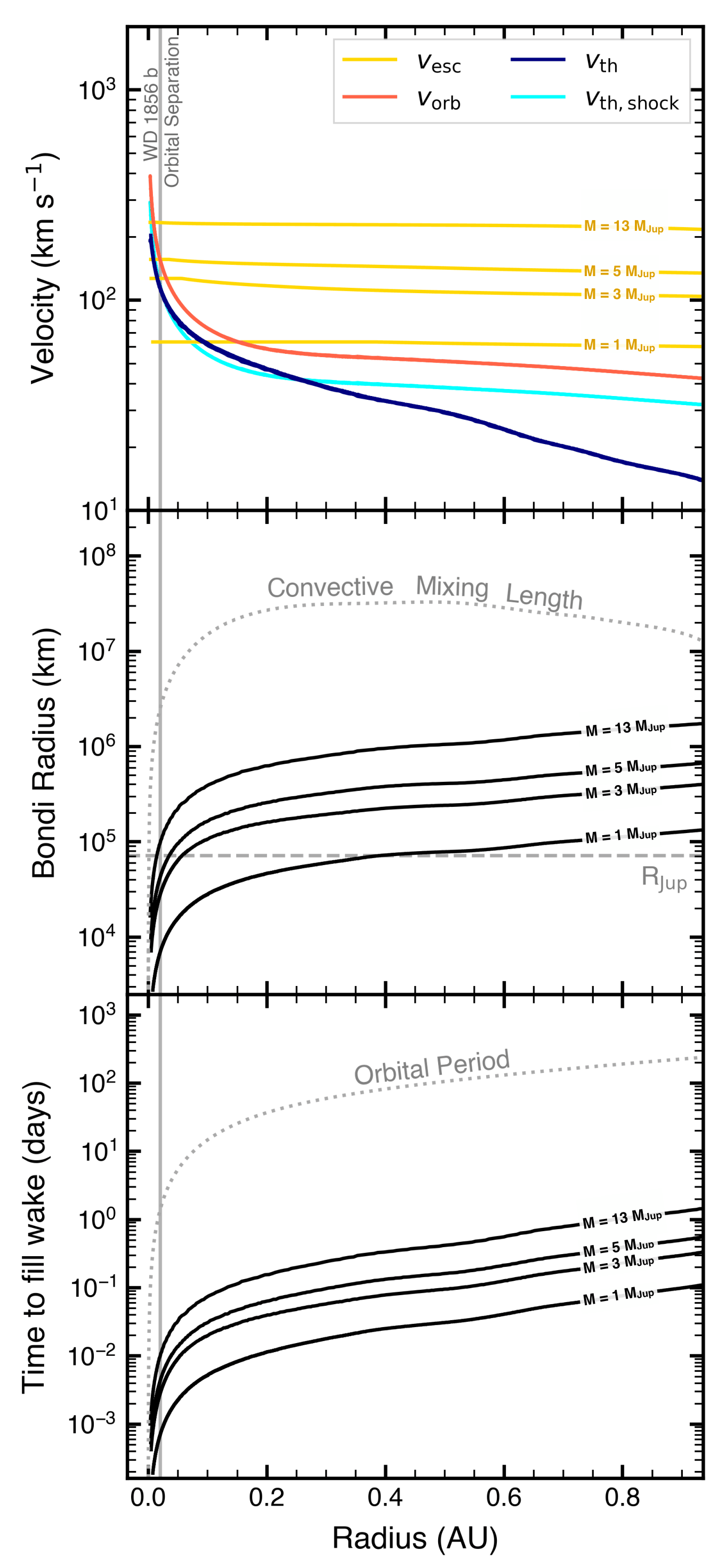}
\caption{Potential complications that may effect the accretion process (see \S\ref{sec:Complications}). The vertical gray line in all figures corresponds to the orbital separation of WD 1856 b. \textit{Top}: Comparison of the planets' orbital velocity $v_{\mathrm{orb}}$, the planets'
escape velocities $v_{\mathrm{esc}}$, the thermal velocity $v_{\mathrm{th}}$ of the surrounding medium, 
and the thermal velocity of the shock front $v_{\mathrm{th,shock}}$ as a function of the radial position within the stellar envelope. \textit{Middle}: Bondi radii $R_{\mathrm{B}}$ of simulated planets relative to the convective mixing length scale $l$. We find that, for each planet, $R_{\mathrm{B}}<<l$ throughout most of the star, meaning we can approximate the planet's surroundings as a uniform medium. We also denote a radius of 1 R$_{\mathrm{Jup}}$ with a horizontal dashed line. \textit{Bottom}: Time \bedit{taken to} fill planet's wake relative to the planet's orbital period. 
\label{Complications}}
\end{figure}

\subsection{Planet Model}
\label{Planet_model}
We treat the inspiralling planet as a point mass within the stellar envelope. To model the planet's orbital path, we use the approach described by \citet{oco2023}, \bedit{available online \citep{oco_zenodo}.} 
This implementation uses {\tt MESA} to model heat deposition from the engulfed planet, which can lead to a drastic expansion of the envelope. See \citet{oco2023} for a discussion of the drag force, orbital decay, and heat deposition. 
As in \citet{oco2023}, we only consider the inspiral phase of engulfment, and ignore the grazing phase, when minimal mass is expected to be accreted. We allow the planet to inspiral until it is destroyed through Roche-lobe overflow (and subsequent tidal disruption) or ram pressure. The radius at which the planet will fill its Roche lobe and be disrupted $a_{\mathrm{dis}}$ is defined as \citep{egg1983}: 
\begin{equation}
a_{\mathrm{dis}} \simeq 2R_{\mathrm{p}} \left(\frac{M_{\mathrm{core}}}{M_{\mathrm{p}}}\right)^{1/3} 
\end{equation}
where $R_{\mathrm{p}}$ and $M_{\mathrm{p}}$ are the radius and mass of the planet, respectively, and $M_{\mathrm{core}}$ is the mass of the stellar core. The condition for planet disruption due to ram pressure is when the ram pressure integrated over the cross-section of the planet approaches the planet's binding energy. This occurs when:
\begin{equation}
\rho_{\mathrm{ext}}\approx\bar{\rho}_{\mathrm{p}}\frac{v_{\mathrm{esc}}^2}{v^2} 
\end{equation}
where $\rho_{\mathrm{ext}}$ is the density of the stellar environment, $\bar{\rho}_{\mathrm{p}}$ is the mean density of the planet, $v_{\mathrm{esc}}$ is the planet's escape velocity, and $v$ is the relative velocity of the planet \citep{jia2018}. 


\subsection{Bondi-Hoyle-Lyttleton Accretion}
\label{Bondi-Hoyle}
We assume the planet accretes mass via Bondi-Hoyle-Lyttleton Accretion \citep[BHLA,][]{1939PCPS...35..405H, 1944MNRAS.104..273B, 1952MNRAS.112..195B}. BHLA assumes that a point particle moves at a constant velocity through a uniform, static gas cloud. It also assumes negligible pressure, such that gas particles follow ballistic orbits, and that there are no external influences, such as magnetic fields. The predicted mass accretion rate $\dot{M}_{\mathrm{BHL}}$ is then defined in terms of the density of the surrounding gas $\rho$ and the local sound speed $c_s$: 
\begin{equation}
\dot{M}_{\mathrm{BHL}}=\frac{4\pi G^2M_{\mathrm{p}}^2\rho}{(c_s^2+v^2)^{3/2}}
\end{equation}
We allow the planet to accrete until the Bondi radius $R_{\mathrm{B}}$, defined as the radius within which material will accrete onto the planet, becomes less than 1 R$_{\mathrm{Jup}}$. Beyond this threshold, the effectiveness of accretion via collision remains poorly understood \citep[see e.g.,][]{orm2024}. Because BHLA assumes a simplified, symmetric accretion flow, it tends to overestimate the amount of accreted material. This has been demonstrated by several previous studies modeling CEE \citep[e.g.,][]{ric2008, ric2012, mac2015, mac2017}. In \S\ref{sec:Complications}, we describe some of the complications that may impact the accretion rate. 
To account for these effects, we multiply $\dot{M}_{\mathrm{BHL}}$ by an efficiency factor \bedit{$\varepsilon$} in our simulations. 

\begin{figure*}
\centering
\includegraphics[width=\textwidth]{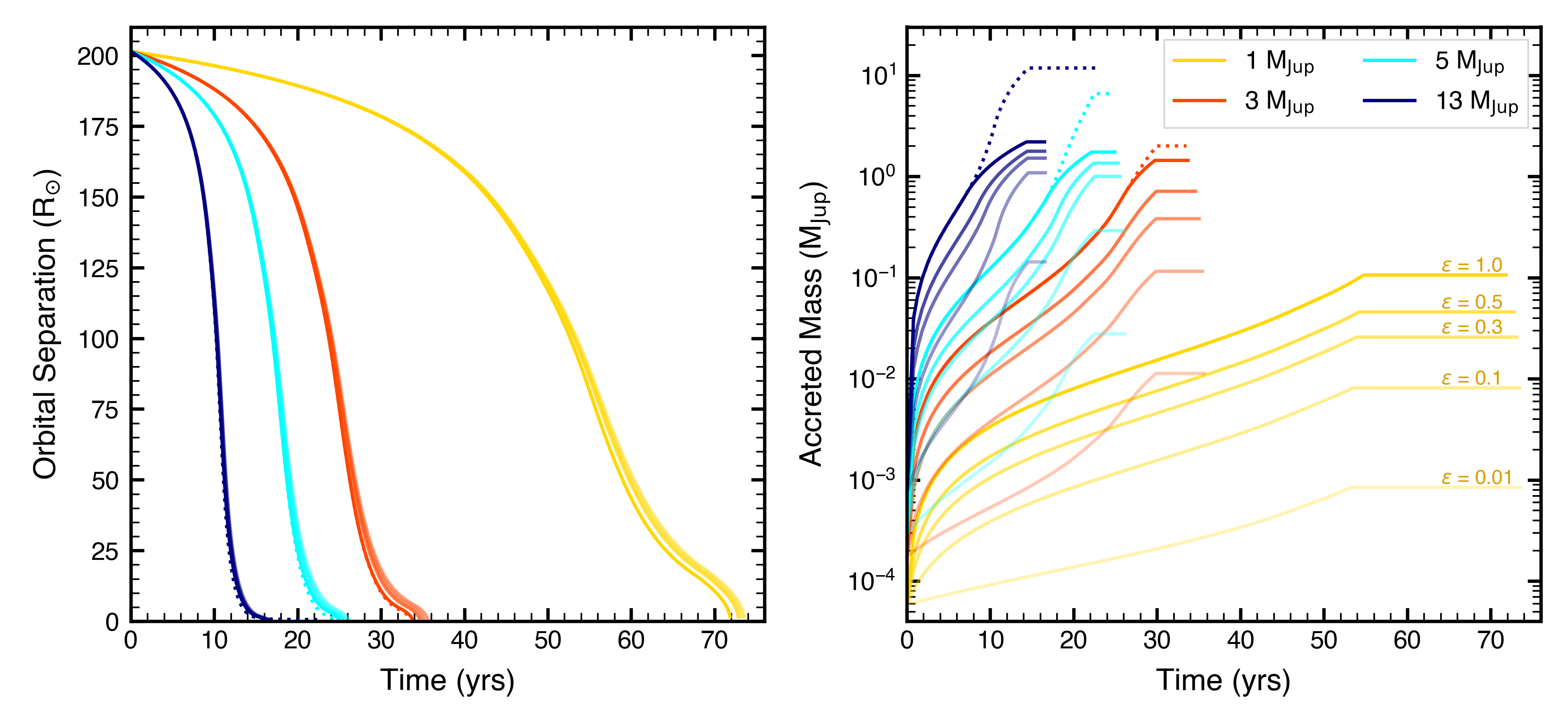}
\caption{\textit{Left}: Orbital separations of inspiralling planets over time. \textit{Right}: Mass accreted onto planets over time. For each initial planet mass, we show results for \bedit{$\varepsilon\in$ [0.01, 0.1, 0.3, 0.5, 1.0]}, with an increasing efficiency denoted by an increasing opacity. 
We also allow the accretion to increase up to 10$\times$ the Eddington limit, denoted by the dotted lines. Accretion stops when $R_B<R_{\mathrm{Jup}}$ (see \S\ref{Bondi-Hoyle}).\label{Accretion}}
\end{figure*}

\subsection{Simulation Description}
\label{Simulation} 
 For each simulation, we inject a planet into the stellar envelope at an initial orbital separation of 0.95 $R_{\star}$. We then simultaneously evolve the planet and stellar models, allowing the planet to accrete mass via BHLA as it inspirals. Accretion terminates when $R_{\mathrm{B}}<1 R_{\mathrm{Jup}}$, and the simulation terminates when the planet has been destroyed by either tidal disruption or ram pressure. This therefore provides an upper limit on how long the planet can spend in the stellar envelope and how much mass it can accrete. We run this simulation for four initial planet masses \bedit{$M_{\mathrm{p,i}}$ $\in$ [1, 3, 5, 13] M$_{\mathrm{Jup}}$ and five values of $\varepsilon$ $\in$ [0.01, 0.1, 0.3, 0.5, 1].} We note that our lower bound on \bedit{$\varepsilon$} is consistent with findings from \citet{ric2012}. In most simulations, we limit the mass accretion rate to be less than the Eddington limit \citep[as calculated for a CEE scenario, ][]{iva2013, cha2018}. However, the Eddington limit can be exceeded, for instance in systems not in hydrostatic equilibrium or with non-uniform accretion. For each planet mass, we therefore also run one simulation with an extended accretion limit of 10$\times$ the Eddington limit. \bedit{The {\tt inlist} and extension files used to run the simulations are available online\footnote{\dataset[doi:10.5281/zenodo.20517223]{https://doi.org/10.5281/zenodo.20517223}}.}

\section{Complications}
\label{sec:Complications}
Because of its simplifying assumptions, BHLA tends to overestimate accretion rates. In this section, we describe some of the complicating effects that may impact the accretion process. While we expect most of the described complications to have a minor impact on the accretion rate, the cumulative impact may be non-negligible. 
We assume that the net effect of these additional complications is to modify the accretion efficiency factor \bedit{$\varepsilon$}.

\subsection{Relative Velocities}
\label{Velocities}
BHLA assumes that the planet accretes from a homogeneous and static medium. Because stellar evolution models predict that AGB stars have low rotational velocities, we model the star as having no rotation, consistent with the BHLA assumption. However, we would expect a stellar envelope to have some non-zero angular momentum. To investigate whether this may have a significant impact on our results, we compare the velocities of the planet and stellar envelope. 
We estimate the rotational velocity of the stellar envelope by evolving the {\tt MESA} model described in \S\ref{Stellar_model}, with the additional assumption of a zero-age MS rotation of rate of $0.5\Omega_{\mathrm{crit}}$, where $\Omega_{\mathrm{crit}}$ is the critical angular velocity \citep{zor2012}. We find that the rotational velocity remains $<$2 km s$^{-1}$ throughout the convective stellar envelope. The rotational velocity then continues to decrease, on average, through the AGB pulsations. In comparison, the engulfed planets' orbital velocities are 1$-$2 orders of magnitude larger throughout the stellar envelope (as shown in the top panel of Figure~\ref{Complications}). 
Our assumption of a non-rotating star should therefore have a negligible effect on our final results. The orbital velocity also exceeds the thermal motion of the surrounding gas at all stellar radii (see Figure~\ref{Complications}), though we note that the difference decreases at smaller radii. 


\begin{table*}[t!]
\centering
\caption{Simulation results. We run our simulation for four initial planet masses \bedit{$M_{\mathrm{p,i}}$ $\in$ [1, 3, 5, 13] M$_{\mathrm{Jup}}$ and five accretion efficiencies $\varepsilon\in$ [0.01, 0.1, 0.3, 0.5, 1.0]}. For each planet mass, we also include one simulation with \bedit{$\varepsilon=1.0$} with an extended upper limit of 10$\times$ the Eddington limit (labeled 1.0+). For each simulation, we provide the total accreted mass $\Delta M$, the fractional increase in the planet's mass $\Delta M/M$, the time before the planet is destroyed $\tau_d$, and the planet's orbital separation at the time of destruction $R_d$.}  \label{tab:Results}
\begin{tabular}{l *{12}{l}}
\toprule
 & \multicolumn{6}{c}{1 M$_{\mathrm{Jup}}$}
 & \multicolumn{6}{c}{3 M$_{\mathrm{Jup}}$} \\
\cmidrule(lr){2-7}
\cmidrule(lr){8-13}
\bedit{$\varepsilon$}
 & 0.01 & 0.1 & 0.3 & 0.5 & 1.0 & 1.0+
 & 0.01 & 0.1 & 0.3 & 0.5 & 1.0 & 1.0+ \\
\midrule
$\Delta M$ (M$_{\mathrm{Jup}}$)  & \bedit{0.001}  & 0.01  & 0.03  & 0.05  & 0.11  & 0.11  & 0.01  & 0.12  & 0.38  & 0.72  & 1.45  & 2.01  \\ 
$\Delta M$/$M$ & \bedit{0.001}  & 0.01  & 0.03  & 0.05  & 0.11  & 0.11  & \bedit{0.004}  & 0.04  & 0.13  & 0.24  & 0.48  & 0.67 \\
$\tau_d$ (yr) & 73.6  & 73.5  & 73.1  & 72.8  & 71.8  & 71.8  & 35.6  & 35.4  & 35.0  & 34.6  & 33.7  & 33.5  \\
$R_d$ (R$_{\odot}$) & 1.69  & 1.69  & 1.69  & 1.67  & 1.64  & 1.64  & 1.18  & 1.14  & 1.12  & 1.08  & 1.02  & 0.98  \\
\hline
\hline
 & \multicolumn{6}{c}{5 M$_{\mathrm{Jup}}$}
 & \multicolumn{6}{c}{13 M$_{\mathrm{Jup}}$} \\
\cmidrule(lr){2-7}
\cmidrule(lr){8-13}
\bedit{$\varepsilon$}
 & 0.01 & 0.1 & 0.3 & 0.5 & 1.0 & 1.0+
 & 0.01 & 0.1 & 0.3 & 0.5 & 1.0 & 1.0+\\
\midrule
$\Delta M$ (M$_{\mathrm{Jup}}$)   & 0.03  & 0.29  & 1.01  & 1.36  & 1.75  & 6.60  & 0.14  & 1.09  & 1.52  & 1.78  & 2.20  & 11.84 \\ 
$\Delta M$/$M$ & 0.01  & 0.06  & 0.20  & 0.27  & 0.35  & 1.32  & 0.01  & 0.08  & 0.12  & 0.14  & 0.17  & 0.91  \\ 
$\tau_d$ (yr) & 26.1  & 25.9  & 25.6  & 25.3  & 24.9  & 24.5  & 16.6  & 16.6  & 16.6  & 16.5  & 16.5  & 22.8  \\ 
$R_d$ (R$_{\odot}$) & 0.97  & 0.96  & 0.93  & 0.91  & 0.88  & 0.74  & 0.72  & 0.70  & 0.68  & 0.69  & 0.67  & 0.58  \\ 
 \bottomrule
\end{tabular}
\end{table*}

\subsection{Convective Mixing}
\label{Turbulence}
Because AGB envelopes are convective, convective motion will create a slowly varying background that the planet moves through. Using the {\tt MESA} stellar model, we estimate that the convective mixing zone reaches a depth of $\approx$0.58 R$_{\odot}$, and that all simulated planets therefore stay in the convective zone throughout their inspiral (see planets' orbital separations at the time of destruction in Table \ref{tab:Results}). To estimate the impact of convective mixing, 
we compare the convective mixing length scale $l$ to the planet's Bondi radius $R_{\mathrm{B}}$. We estimate $l$ using the relation $l=\alpha_{\mathrm{MLT}} H_{\mathrm{P}}$, where $H_{\mathrm{P}}$ is the scale height. Assuming hydrostatic equilibrium, we can find $H_{\mathrm{P}}$ :
\begin{equation}
    H_{\mathrm{P}}=-\frac{dr}{d\mathrm{ln}P}=\frac{P}{\rho g}
\end{equation} 
where the local pressure is $P$, gas density is $\rho$, and gravitational acceleration is $g$.
Adopting $\alpha_{\mathrm{MLT}}=2$, we obtain
\begin{equation}
l=\frac{2P}{\rho g} .
\end{equation}
The Bondi radius $R_{\mathrm{B}}$ is defined as the radius within which material will accrete onto the planet: 
\begin{equation}
R_{\mathrm{B}}=\frac{2GM_{\mathrm{p}}}{v^2+c_s^2}
\end{equation} 
The predicted values of $l$ and $R_{\mathrm{B}}$ throughout the star's envelope are shown in the middle panel of Figure~\ref{Complications}. We find that $l$ exceeds $R_{\mathrm{B}}$ throughout the stellar interior, and is often over one order of magnitude larger. 
For instance, at orbital separations greater than 0.0204 au (the orbital separation of WD 1856 b),  $l$ is 7$-$44 times larger than the 13 M$_{\mathrm{Jup}}$'s $R_{\mathrm{B}}$. Because the convection length scales are significantly larger than the planet's accretion length scales, we can approximate the planet's surrounding medium as relatively uniform, meaning we do not expect these spatial variations to drive any large deviations from the simplified BHLA calculation.

Additionally, if convective mixing changes the background on short timescales, then the background can no longer be considered static. To evaluate how important this effect is, we compare the convection timescale $\tau_{\mathrm{conv}}$ to the simulated planet's orbital period. Defining $\tau_{\mathrm{conv}}$ as $l$/v$_{\mathrm{conv}}$, where v$_{\mathrm{conv}}$ is the local convective velocity, we find that $\tau_{\mathrm{conv}}$ tends to be the same order of magnitude as the orbital period throughout the convective envelope, and exceeds the orbital period for separations $\lesssim127$ R$_{\odot}$ ($0.59$ au). At close-in separations, $\tau_{\mathrm{conv}}$ is up 5.7 times larger than the planet's orbital period. This suggests that convective mixing may drive non-negligible changes in the background during a planet's inspiral. These largely stochastic fluctuations may serve to either increase or decrease the amount of mass accreted, and further work is needed to determine the net impact of turbulence on engulfed planets.

\subsection{Accretion Wake}
\label{Accretion Wake}
We expect the planet to form a wake as it orbits within the stellar envelope. If the surrounding gas does not fill the wake before the planet completes a full orbit, then there will not be as much surrounding material for the planet to accrete. To estimate how long it takes to fill the wake, we calculate the thermal velocity of the gas $v_{\mathrm{th}}$ from the star's temperature $T$ and mean molecular weight $m$:
\begin{equation}
v_{\mathrm{th}}=\sqrt{\frac{3k_{\mathrm{B}}T}{ m}}
\end{equation}
We use our stellar {\tt MESA} model to estimate $T$ and $m$ throughout the stellar envelope. Assuming the size of the wake is $\approx2 R_{\mathrm{B}}$, we estimate the average timescale for filling the wake to be 0.04 $-$ 0.51 days for our initial planet mass range. As shown in the bottom panel of Figure~\ref{Complications}, the orbital period remains approximately two orders of magnitude larger throughout the convective envelope. We therefore conclude that the wake will have sufficient time to fill throughout the planet's inspiral, and that this will not impact the final accretion rate.

\begin{figure}
\centering
\includegraphics[width=\columnwidth]{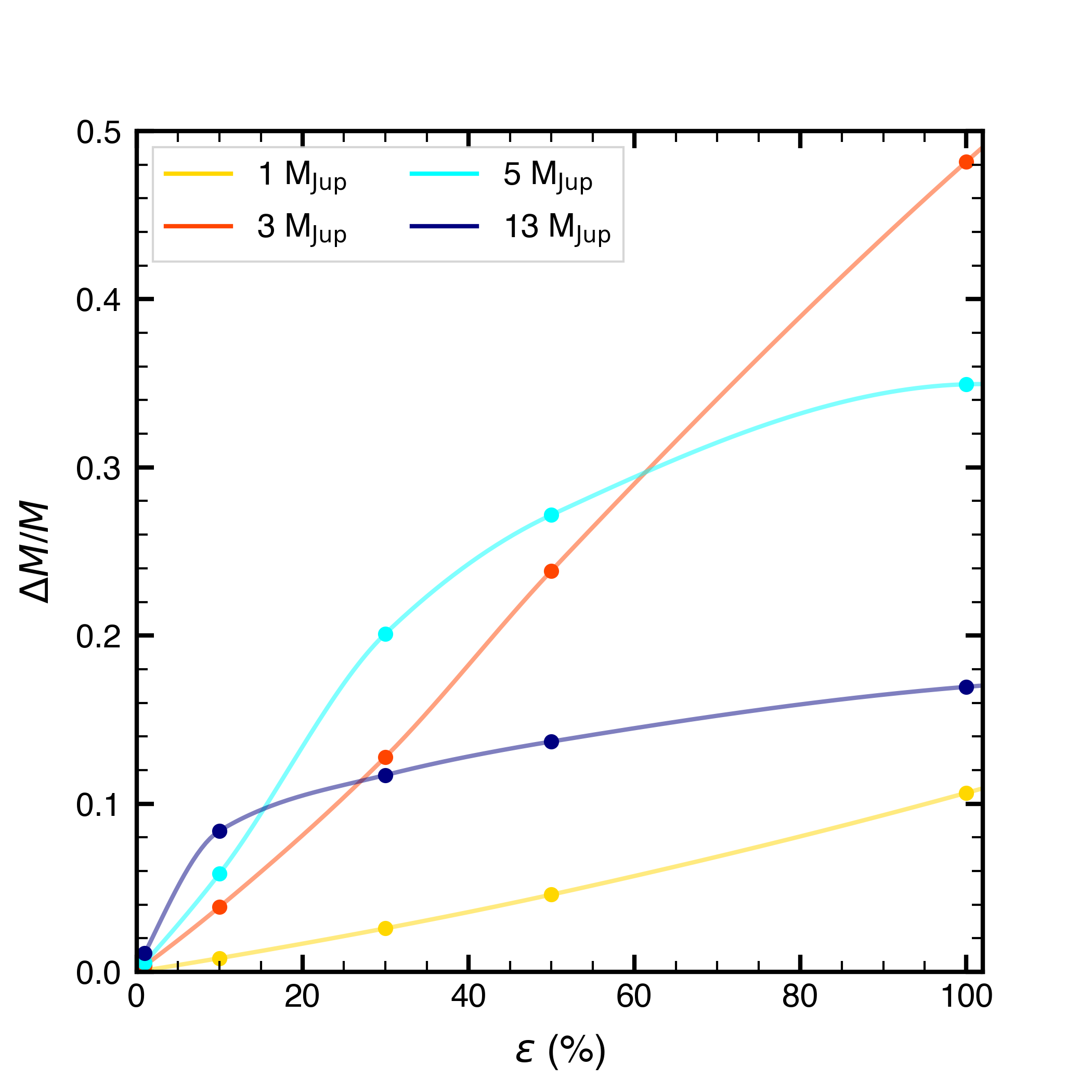}
\caption{Fractional increase in mass as a function of accretion efficiency \bedit{$\varepsilon$}. Each line in an interpolation of the fractional increase for a given initial mass, calculated using monotonic cubic splines. For many values of \bedit{$\varepsilon$}, the planet experiences a significant increase in mass. The 3 M$_{\mathrm{Jup}}$ planet exhibits the largest increase for \bedit{$\varepsilon=1.0$}, with $\Delta M$/$M=$ 0.48. \label{Efficiency}}
\end{figure}

\subsection{Magnetic Fields}
\label{Magnetic_fields}
BHLA assumes that there are no external forces. However, if there is a significant stellar magnetic field, this assumption may be invalid. To investigate the importance of magnetic fields, we first compare the relative strengths of the stellar and planetary magnetic fields. 
The magnetic fields of AGB stars are often approximated to be several to tens of Gauss at the surface \citep[e.g.,][]{sea2000, gar2001, pas2020}, and observations estimate a wide range of values \citep[see reviews by e.g.,][]{vle2019, kon2024}. 
\citet{ohl2016} modeled how a stellar magnetic field would then evolve in a CEE scenario. Starting with a 2 M$_{\odot}$ red giant with a magnetic field strength of 10$^{-6}$ G at the stellar surface, they simulated an interaction with an engulfed 1 M$_{\odot}$ companion. They evolved the magnetic field as the companion spirals inward, and found the magnetic field strength increased up to $\sim$10$^3$ G after 120 days. An engulfed planet would likely have a much smaller impact on the host star's magnetic field, though this remains largely unstudied. 
In comparison, Jupiter's magnetic field ranges up to $\sim$10 G \citep[e.g.,][]{con1993, con2018, liv2024}. 
Jupiter analogues that are closer to their star or are more massive can have even larger magnetic fields. For instance, surface magnetic fields in the range of 20 $-$ 120 G have been measured for hot Jupiters \citep{2019ESS.....431909C}, and it's predicted that the magnetic field could increase to $\sim$250 G for a $\approx$10 M$_{\mathrm{Jup}}$ hot Jupiter \citep{2017ApJ...849L..12Y}. Therefore, assuming an AGB magnetic field up to several Gauss, we predict that an engulfed giant planet's magnetic field will be locally dominant and that
the stellar magnetic field will therefore have a negligible effect on the accretion process. However, if the stellar magnetic field is enhanced during CEE, as predicted for stellar binaries, than this may no longer be a valid assumption. Future work could investigate potential effects of a non-negligible stellar magnetic field on planetary accretion (see \S\ref{Future work}). 

The planetary magnetic field, however, may introduce additional complications. For instance, interactions between atmospheric winds and the planet's magnetic field can drive ohmic heating if the atmosphere is partially ionized, as has been proposed for hot Jupiter planets \citep[e.g.,][]{bat2010, per2010, hua2012}. This could potentially contribute $>1\%$ of the stellar insolation power \citep{per2010}, particularly in deeper atmospheric layers, and drive significant radius inflation \citep[e.g., ][]{bat2010, per2010, Batygin2025}. Future work could include such effects in the accretion model (see \S\ref{Future work}). 

\subsection{Shocks}
As infalling material approaches freefall velocities, it may drive shocks that expel some of the accreting gas. This has been observed on several accreting planetary-mass objects, including PDS 70 b \citep{aoy2019, haf2019, has2020}, Delorme 1(AB)b \citep{bet2022, rin2023}, and TWA 27B \citep{luh2023, 2024ApJ...964...70M}.
The fractional amount of energy that leaves the system can be quantified by 
\begin{equation}
\eta^{phy}=\frac{\dot{E}(r_{\mathrm{max}})-\dot{E}(r_{\mathrm{ps}})}{\dot{E}(r_{\mathrm{max}})}
\end{equation}
where $\dot{E}(r)$ is the total energy flux measured at some radius $r$, $r_{\mathrm{max}}$ is a proxy for the accretion radius, and $r_{\mathrm{ps}}$ is the post-shock radius \citep{2017ApJ...836..221M}. While some hydrodynamical simulations predict $\eta^{phy}$ close to 1 \citep[$\gtrsim$0.97,][]{2017ApJ...836..221M,2019ApJ...881..144M}, simulations including hydrogen dissociation and ionization find that $\eta^{phy}$ can be lower \citep[$\gtrsim$0.65,][]{Chen__2022} in systems with higher accretion rates. This may therefore have a significant impact on the overall accretion rate. 

As the planet inspirals through the stellar envelope, it may also create a shock front. If the thermal velocity of the shock front is greater than the planet's escape velocity, then this will reduce the amount of material accreting onto the planet from the direction of the shock. To estimate the thermal velocity of the shock front, we first estimate its temperature $T_{\mathrm{s}}$ using the Rankine–Hugoniot shock jump conditions. Assuming an ideal gas, we calculate: 
\begin{equation}
T_{\mathrm{s}} = \frac{3}{16k_{\mathrm{B}}} m u_{\mathrm{s}}^2
\end{equation}
where $u_{\mathrm{s}}$ is the velocity of the shock front. We estimate $T_{\mathrm{s}}\sim10^4-10^6$ K throughout most of the stellar envelope. As shown in the top panel of Figure~\ref{Complications}, the corresponding thermal velocity is less than the planet's escape velocity throughout most of the star's interior. 
We therefore conclude that a shock front would not prevent a significant amount of mass from accreting onto the planet. However, because the difference between escape velocity and thermal velocity can be relatively small (particularly for less massive planets), more detailed calculations may be needed to confirm this.

\subsection{Planetary Heating and Winds} 
Planetary heating and thermally driven winds could also have a significant effect on the accretion rate. Based on the virial theorem, we expect around half of the infalling material's potential energy to be deposited as thermal energy. Therefore, as the planet accretes mass, its atmosphere will heat up and expand. If the accretion timescales are shorter than the planet's cooling timescales, then this will eventually prevent further accretion. Depending on when this stage is reached, accretion may be prematurely stopped, reducing the amount of accreted material. The planet can also lose mass via evaporation or thermally driven winds. This is characterized by the Jean's parameter $\lambda$, which is the ratio between gravitational potential energy and thermal energy at the top of the atmosphere. $\lambda$ is defined as $\lambda = \frac{GM_{\mathrm{p}}m}{rk_{\mathrm{B}}T(r)}$, where $r$ is the distance from the planet's center and $T(r)$ is the temperature at that distance. When $\lambda \gg10$, the atmosphere is tightly bound, preventing escape. However, when $\lambda\sim10$, thermally driven escape occurs. 
Because we only model the engulfed planet as a point source, we do not account for additional heating, atmospheric inflation, or evaporation. However, we discuss in \S\ref{Evaporation} whether we expect the planets to completely evaporate, and we discuss how future work could account for these effects in \S\ref{Future work}.

\begin{table}[t]
\centering
\caption{Mass fractions of carbon, nitrogen, and oxygen in the 3 M$_{\mathrm{Jup}}$ planet's atmosphere for each value of \bedit{$\varepsilon$}. \bedit{Mass fractions for the initial planet that has not undergone CEE are labeled as 0.0, and results} for \bedit{$\varepsilon=1$} with an extended upper limit of 10$\times$ the Eddington limit are labeled as 1.0$+$. These mass fractions are described in more detail in \S\ref{Composition2}. } \label{tab:Abundances}
\centering
{\footnotesize
\begin{tabular}{rcccccccc}

\toprule
\bedit{$\varepsilon $} & 0.0 & 0.01 & 0.1 & 0.3 & 0.5 & 1.0 & 1.0$+$ \\
\cmidrule(lr){2-8}
$X_{\mathrm{C}}$ & 0.0066 & 0.0066 & 0.0063 & 0.0059 &0.0055 & 0.0050 & 0.0047 \\
$X_{\mathrm{N}}$ & 0.0025 & 0.0025 & 0.0024 & 0.0024 & 0.0023 & 0.0022 & 0.0022 \\
$X_{\mathrm{O}}$ & 0.0184 & 0.0183 & 0.0178 & 0.0168 & 0.0159 & 0.0146 & 0.0139 \\
\bottomrule

\end{tabular}}

\end{table}

\subsection{Post-CEE Mass Loss}
\label{Mass loss}
Assuming the stellar envelope is successfully ejected during the CEE phase, the star will then become a hot subdwarf that emits strong EUV radiation. 
This may drive photoevaporative mass loss on the planet. \citet{lag2021} previously estimated this effect for the WD 1856 system, and found that WD 1856 b would have experienced negligible mass loss. We repeat this calculation for our range of planet masses. Assuming the planets are at an orbital separation of 0.0204 au, we model the mass loss due to hydrodynamic escape in both the recombination limited and energy limited regimes as the WD host cools over 10 Gyr. As in \citet{lag2021}, we use the energy-limited mass loss rate for fluxes below 10$^4$ erg cm$^{-2}$s$^{-1}$ \citep{mur2009}:
\begin{equation}
\dot{M} = \frac{\varepsilon_{\mathrm{ML}}\pi F_{\mathrm{EUV}}R_{\mathrm{p}}^3}{GM_{\mathrm{p}}K(\xi)}
\end{equation}
where $F_{\mathrm{EUV}}$ is the flux in the EUV and \bedit{$\varepsilon_{\mathrm{ML}}$} is an efficiency factor set to 0.3 (\citealt{mur2009, owe2016}; however, the efficiency will change with planet mass and WD cooling age; \citealt{Gallo2024}). $K(\xi=R_{\mathrm{L_p}}/R_{\mathrm{p)}}$ is a correction term that accounts for the planet's Roche-lobe $R_{\mathrm{L_p}}$ and radius being on similar scales, which can increase the rate of mass loss \citep{erk2007}. We model the WD's EUV flux as a function of temperature by assuming it emits as a ideal blackbody. We estimate the following fraction of its bolometric luminosity is emitted in the EUV \citep[e.g.,][]{bec2025}: 
\begin{equation}
\frac{L_{\mathrm{XUV}}}{L_{\mathrm{bol}}} = \frac{\int_{\mathrm{XUV}}d \nu  u_{\nu}(\nu,T)}{\int_{0}^{\infty}d \nu  u_{\nu}(\nu,T)}
\end{equation}
where $\nu$ is the frequency and $u_{\nu}(\nu,T)$ is the Planck spectrum. We consider the EUV to include wavelengths between 1 $-$ 1200 \AA. For the least massive planet in our simulations (1 M$_{\mathrm{Jup}}$), which is also the most prone to mass loss, we find that the planet loses $<0.003\%$ of its mass during this post-CEE evaporation. This is also likely an overestimate, given that \bedit{$\varepsilon_{\mathrm{ML}}$} is expected to decrease as the system ages \citep{Gallo2024}. We therefore consider photoevaporative mass loss a negligible effect. 

Additionally, for $\sim$10,000 years after the AGB phase, the planet is embedded in a planetary nebula, characterized in part by high-velocity winds and high stellar effective temperatures. Using the equations described above and the post-AGB evolutionary models from \citet{mil2016}, we expect the planet to lose minimal mass. However, the planet may reach blow-off conditions (as suggested by \citet{vil2007}) when $\lambda\lesssim1.5-3$ \citep[e.g.,][]{opi1963}. In such a scenario, the planet may lose most of its atmosphere, 
therefore removing any signatures of CEE accretion. Further work is needed to explore how this phase would impact close-in giant planets.

\begin{figure*}
\centering
\includegraphics[width=\textwidth]{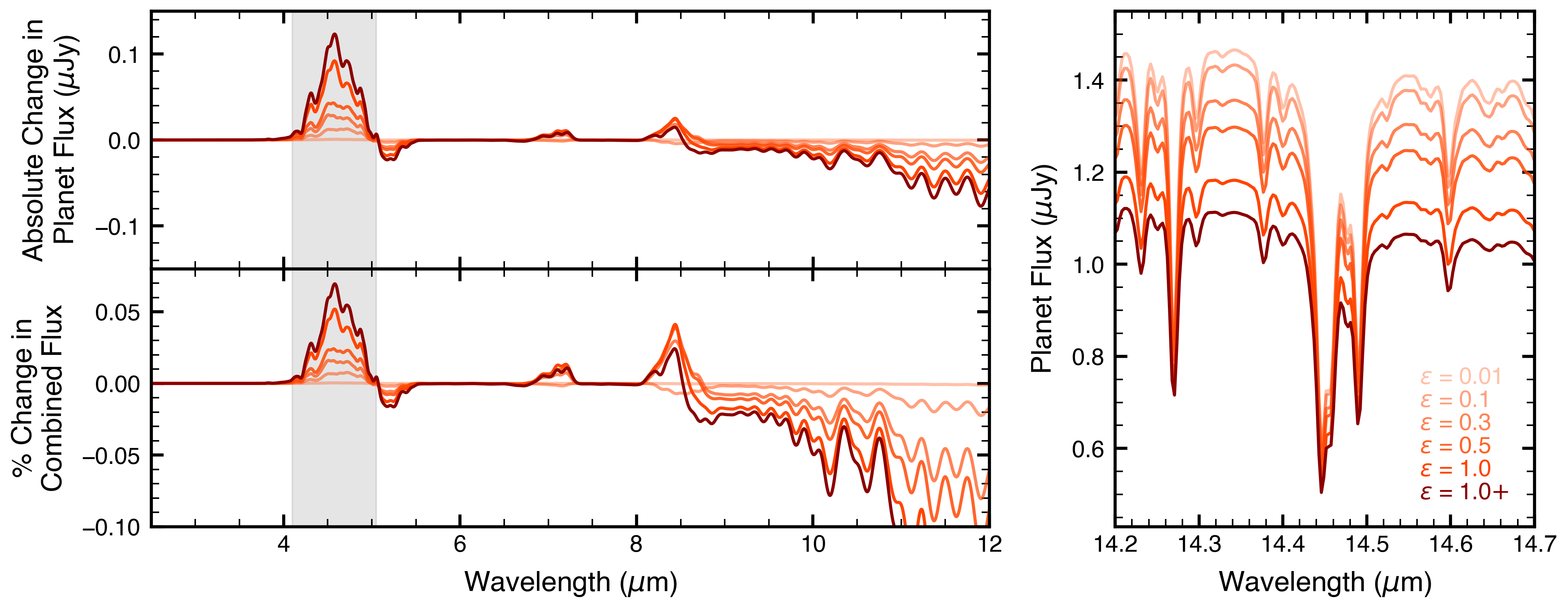}
\caption{\textit{Top Left}: Absolute change in emission of a hypothetical 3 M$_{\mathrm{Jup}}$ planet in the WD 1856 system ($d=$ 24.8 pc). We show results for \bedit{$\varepsilon\in$ [0.01, 0.1, 0.3, 0.5, 1.0]}, with an increasing efficiency denoted by an increasing opacity and a spectral resolution of 200. We also include results for \bedit{$\varepsilon=1$} with an extended upper limit of 10$\times$ the Eddington limit (labeled 1.0$+$). We observe a significant increase in flux in the 4.10$-$5.05 $\mu$m opacity window. We therefore use this feature (denoted by the gray shaded region) to investigate how the CEE signal changes as a function of temperature (see Figure~\ref{Temperature}). \textit{Bottom Left}: Percent change in combined planet and white dwarf spectrum. For an unresolved 3 M$_{\mathrm{Jup}}$ planet, we expect an increase of up to $\sim$0.07\% in the 4.10$-$5.05 $\mu$m opacity window. \textit{Right}: NH$_{\mathrm{3}}$ spectral lines with a spectral resolution of 10000, predicted for different values of \bedit{$\varepsilon$}. If a planet has undergone CEE, we expect the amplitude of metal lines to significantly change. In the \bedit{$\varepsilon=1.0+$} scenario, the depth of the NH$_{\mathrm{3}}$ line is just $\sim$60\% of the predicted depth for a planet that has not undergone CEE.\label{Composition}}
\end{figure*}

\section{Results}
\label{Results}

\subsection{Mass Accretion}
\label{Mass_accretion}
Using the simulations described in \S\ref{Model}, we calculate how much mass a planet accretes as a function of its initial mass and \bedit{$\varepsilon$}. We show each planet's orbital separation and cumulative mass throughout its inspiral in Figure~\ref{Accretion}, and report the accretion results in Table~\ref{tab:Results}. As expected, the amount of accreted mass increases with both the planet's starting mass and \bedit{$\varepsilon$}. \bedit{If we allow super-Eddington accretion (which can occur, for instance, when accretion is non-uniform), we find that higher mass planets can approximately double their mass. For an efficiency factor of 1.0 and Eddington-limited accretion,} the simulated planets accrete between 0.11$-$2.20 M$_{\mathrm{Jup}}$, corresponding to 11$-$48$\%$ of their initial mass. The 3 M$_{\mathrm{Jup}}$ planet exhibits the largest fractional increase in mass---48\% assuming an efficiency factor of 1.0. We therefore adopt the 3 M$_{\mathrm{Jup}}$ planet as our fiducial case for the following analysis in \S\ref{Composition2}, in which we investigate potential compositional changes. Lower values of \bedit{$\varepsilon$} lead to less overall accretion. For \bedit{$\varepsilon =0.01$}, the simulated planets accrete between 0.001$-$0.14 M$_{\mathrm{Jup}}$, corresponding to just 0.1$-$1.1\% of their initial mass. We show the fractional increase in mass as a function of \bedit{$\varepsilon$} in Figure~\ref{Efficiency}. 


As previously noted, these results assume the planet accretes material until it is destroyed, providing an upper limit on how much mass can be accreted. If the stellar envelope is ejected before this point, allowing the planet to survive, then the planet will spend less time in the envelope and will therefore accrete less mass. In particular, if the envelope is ejected when the planet reaches the orbital separation of WD 1856 b, we do not predict a significant change in the amount of mass accreted. For the 1 and 3 M$_{\mathrm{Jup}}$ planets, we expect no change; for the 5 M$_{\mathrm{Jup}}$ planet we expect a decrease of 0.15 M$_{\mathrm{Jup}}$ in the super-Eddington scenario; and for the 13 M$_{\mathrm{Jup}}$ planet we expect a maximum decrease of 0.09 M$_{\mathrm{Jup}}$ in the Eddington-limited scenarios and a decrease of 2.34 M$_{\mathrm{Jup}}$ in the super-Eddington scenario. Because we focus only on the 3 M$_{\mathrm{Jup}}$ planet for our investigation in \S\ref{Composition2}, we would not expect our following results to change if we were to specifically calculate the mass accretion for a WD 1856 b-like planet. 

\subsection{Compositional Changes}
\label{Composition2}
Accretion from the stellar envelope will alter a planet's atmospheric composition. While the composition of the AGB envelope will change over time \citep[e.g.,][]{ven2015, ven2016_2}, we assume here that the planet is engulfed at the tip of the first thermal pulse, as described in \S\ref{Stellar_model}. At this stage of its evolution, the AGB stellar envelope will likely have lower carbon, nitrogen, and oxygen abundances relative to the giant planet. As the planet accretes material from the star, it will therefore become enriched in hydrogen and helium. To investigate this potential observational signature, we simulate emission spectra for a 3 M$_{\mathrm{Jup}}$ planet that has accreted from a stellar envelope. We assume the planet has an initial metallicity of [M/H] = +0.3 and C/O = 1.0 (solar), and obtain atmospheric abundances using reference chemistry tables from \cite{lupu_2021}. This provides initial carbon, nitrogen, and oxygen mass fractions of $X_{\mathrm{C}}= $ 0.0066, $X_{\mathrm{N}}= $ 0.0025, and $X_{\mathrm{O}} = $ 0.0184, respectively. We then account for mass accreted from the stellar envelope, assumed to have a composition calculated using our {\tt MESA} model ($X_{\mathrm{C}}= $ 0.0016, $X_{\mathrm{N}}= $ 0.0017, and $X_{\mathrm{O}}= $ 0.0066). The \bedit{initial and} final mass fractions of carbon, nitrogen, and oxygen in the planet's atmosphere are given in Table~\ref{tab:Abundances} as a function of \bedit{$\varepsilon$}. 

\bedit{To} generate the simulated spectra, we use \texttt{PICASO} \bedit{version 4.0 \citep{picaso_zenodo}}, an open source Python package that computes atmospheric models in radiative-convective equilibrium, along with thermal emission, reflected light, and transmission spectra \citep{batalha2019,mukherjee2023,mang2026}. For this work, we assume the atmosphere is in chemical equilibrium. For each value of \bedit{$\varepsilon$}, we simulate the spectrum assuming a temperature of 186 K \citep[\bedit{consistent with the temperature of WD 1856 b, as calculated with a blackbody fit to the observed infrared excess, }][]{lim2025}. We modify the abundances of carbon-, nitrogen-, and oxygen-bearing species relative to the fiducial 186 K model, and construct corresponding P-T profiles by applying a uniform temperature offset to a fiducial profile. The offset is determined by iteratively adjusting the temperature profile and recomputing the total emergent flux until the target effective temperature is reached. This approach preserves the shape of the original P-T profile while ensuring consistency in effective temperature across models with post-processed abundances.
The resulting emission spectra for a hypothetical 3 M$_{\mathrm{Jup}}$ planet in the WD 1856 system (with a distance $d=$ 24.8 pc) are shown in Figure ~\ref{Composition}. 

We find that accretion reduces the amplitude of metal lines. \bedit{For instance, we expect a planet that has not undergone CEE to have up to $\sim$60\% deeper NH$_3$ spectral lines than a planet that has undergone CEE. We demonstrate how NH$_3$ varies as a function of accretion in Figure~\ref{Composition}. However, } 
observing such effects on individual spectral lines requires high spectral resolution and long exposure times to obtain sufficient signal-to-noise ratios, making them difficult to observe. However, bulk metallicity effects may be more readily characterized. We expect that accretion will impact the opacity windows in the planet's atmosphere. With fewer metals, the window extends to deeper, hotter layers in the atmosphere, resulting in more thermal emission. Notably, in the \bedit{4.10$-$5.05} $\mu$m regime, the thermal emission increases by up to \bedit{7.8 $\times$ 10$^{-21}$ W m$^{-2}$ in the most extreme case}, resulting in a \bedit{9.0}\% increase in planet's spectrum and a \bedit{0.03\% increase} in the combined WD and planet spectrum. Additionally, the planet contributes more flux at redder wavelengths, amplifying the fractional change in emission. The fractional change at $\gtrsim$10 $\mu$m is therefore comparable to the fractional change predicted at 4$-$5 $\mu$m, making it particularly favorable to search for discrepancies at redder wavelengths. \bedit{Given that BHLA is known to overestimate the amount of accretion, we note that the $\varepsilon=1.0$ and $1.0+$ scenarios are upper bounds on what we would expect to observe. As $\varepsilon$ decreases, the hypothetical planet accretes less material and the signatures become smaller. For instance, for $\varepsilon\in$ [0.01, 0.1, 0.3, 0.5], planet emission increases by just 0.1$-$3.6\%. In the $\varepsilon=0.01$ scenario, a planet that has undergone CEE is nearly indistinguishable from a planet that has not undergone CEE. We provide the absolute and fractional change in thermal emission for each scenario in Table~\ref{tab:Composition}.} 

These results are for a cold (186 K), distant ($d=$ 24.8 pc) world \bedit{similar to WD 1856 b.} For a hotter planet, we expect the emission spectrum to exhibit larger changes \bedit{between the CEE and non-CEE scenarios}. To investigate the importance of temperature, we generate simulated emission spectra for a range of effective temperatures \bedit{$T_{\mathrm{eff}}\in$ [200, 250, 300, 350, 400] K,} using the same methods previously described. We then calculate the absolute change in planet flux in the 4.10$-$5.05 $\mu$m window (corresponding to the shaded region in Figure~\ref{Composition}) for each spectrum. Our results are shown in Figure~\ref{Temperature} \bedit{and provided in Table~\ref{tab:Composition}.} We find that the expected change in emission increases as a function of the planet's temperature, with a 400 K planet exhibiting an increase of \bedit{up to $1.7\times10^{-19}$ W m$^{-2}$ assuming $\varepsilon=1.0$ and an extended upper limit of 10$\times$ the Eddington limit.} Such a discrepancy could be observable with present-day capabilities. We discuss prospects for observing signatures of CEE in \S\ref{JWST}.




\begin{table*}[t]
\centering
\caption{\bedit{Thermal emission signatures. We generate emission spectra for a 3 M$_{\mathrm{Jup}}$ planet assuming a range of five accretion efficiencies $\varepsilon\in$ [0.01, 0.1, 0.3, 0.5, 1.0] and six effective temperatures $T_{\mathrm{eff}}\in$ [186, 200, 250, 300, 350, 400] K. As before, we also include a set of simulations with \bedit{$\varepsilon=1$} with an extended upper limit of 10$\times$ the Eddington limit, labeled as 1.0$+$. For each scenario, we provide the absolute change in planet emission between 4.10$-$5.05 $\mu$m ($\Delta F_{\mathrm{p}}$, in units of W m$^{-2}$), fractional increase in planet emission relative to a planet that has not undergone CEE ($\Delta F_{\mathrm{p}}/F_{\mathrm{0}}$), and fractional increase in the total planet and white dwarf emission ($\Delta F_{\mathrm{p}}/F_{\mathrm{t}}$). } } \label{tab:Composition}
\hspace*{-2cm}
{
\begin{tabular}{lrcccccc}
\cmidrule[0.08em]{1-8}

$T$ = 186 K  & $\varepsilon$ & 0.01 & 0.1 & 0.3 & 0.5 & 1.0 & 1.0+ \\
\cline{3-8}
  & $\Delta F_{\mathrm{p}}$   & 4.9 $\times$ 10$^{-23}$  & 9.3 $\times$ 10$^{-22}$  & 2.0 $\times$ 10$^{-21}$  & 3.1 $\times$ 10$^{-21}$  & 5.8 $\times$ 10$^{-21}$  & 7.8 $\times$ 10$^{-21}$\\
 &   $\Delta F_{\mathrm{p}}/F_{\mathrm{0}}$  & 5.7 $\times$ 10$^{-4}$  & 1.1 $\times$ 10$^{-2}$  & 2.3 $\times$ 10$^{-2}$  & 3.6 $\times$ 10$^{-2}$  & 6.7 $\times$ 10$^{-2}$  & 9.0 $\times$ 10$^{-2}$ \\
 & $\Delta F_{\mathrm{p}}/F_{\mathrm{t}}$   & 2.0 $\times$ 10$^{-6}$ & 3.8 $\times$ 10$^{-5}$ & 8.2 $\times$ 10$^{-5}$ & 1.3 $\times$ 10$^{-4}$ & 2.3 $\times$ 10$^{-4}$ & 3.1 $\times$ 10$^{-4}$\\
\cline{1-8}
\\[-3ex]
\cline{1-8}
$T$ = 200 K  & $\varepsilon$ & 0.01 & 0.1 & 0.3 & 0.5 & 1.0 & 1.0+ \\
\cline{3-8}
  & $\Delta F_{\mathrm{p}}$   & 6.8 $\times$ 10$^{-23}$  & 4.3 $\times$ 10$^{-22}$  & 3.3 $\times$ 10$^{-21}$  & 4.5 $\times$ 10$^{-21}$  & 10.0 $\times$ 10$^{-21}$  & 1.2 $\times$ 10$^{-20}$\\
 &   $\Delta F_{\mathrm{p}}/F_{\mathrm{0}}$  & 5.2 $\times$ 10$^{-4}$  & 3.3 $\times$ 10$^{-3}$  & 2.5 $\times$ 10$^{-2}$  & 3.4 $\times$ 10$^{-2}$  & 7.6 $\times$ 10$^{-2}$  & 9.0 $\times$ 10$^{-2}$ \\
 & $\Delta F_{\mathrm{p}}/F_{\mathrm{t}}$   & 2.7 $\times$ 10$^{-6}$ & 1.7 $\times$ 10$^{-5}$ & 1.3 $\times$ 10$^{-4}$ & 1.8 $\times$ 10$^{-4}$ & 4.0 $\times$ 10$^{-4}$ & 4.7 $\times$ 10$^{-4}$\\
\cline{1-8}
\\[-3ex]
\cline{1-8}
$T$ = 250 K  & $\varepsilon$ & 0.01 & 0.1 & 0.3 & 0.5 & 1.0 & 1.0+ \\
\cline{3-8}
  & $\Delta F_{\mathrm{p}}$   & 2.1 $\times$ 10$^{-22}$  & 7.0 $\times$ 10$^{-21}$  & 1.4 $\times$ 10$^{-20}$  & 1.8 $\times$ 10$^{-20}$  & 3.3 $\times$ 10$^{-20}$  & 3.8 $\times$ 10$^{-20}$\\
 &   $\Delta F_{\mathrm{p}}/F_{\mathrm{0}}$  & 4.3 $\times$ 10$^{-4}$  & 1.5 $\times$ 10$^{-2}$  & 3.0 $\times$ 10$^{-2}$  & 3.9 $\times$ 10$^{-2}$  & 7.0 $\times$ 10$^{-2}$  & 8.1 $\times$ 10$^{-2}$ \\
 & $\Delta F_{\mathrm{p}}/F_{\mathrm{t}}$   & 8.2 $\times$ 10$^{-6}$ & 2.8 $\times$ 10$^{-4}$ & 5.7 $\times$ 10$^{-4}$ & 7.3 $\times$ 10$^{-4}$ & 1.3 $\times$ 10$^{-3}$ & 1.5 $\times$ 10$^{-3}$\\
\cline{1-8}
\\[-3ex]
\cline{1-8}
$T$ = 300 K  & $\varepsilon$ & 0.01 & 0.1 & 0.3 & 0.5 & 1.0 & 1.0+ \\
\cline{3-8}
  & $\Delta F_{\mathrm{p}}$   & 3.7 $\times$ 10$^{-22}$  & 1.5 $\times$ 10$^{-20}$  & 2.2 $\times$ 10$^{-20}$  & 3.5 $\times$ 10$^{-20}$  & 5.4 $\times$ 10$^{-20}$  & 6.6 $\times$ 10$^{-20}$\\
 &   $\Delta F_{\mathrm{p}}/F_{\mathrm{0}}$  & 2.9 $\times$ 10$^{-4}$  & 1.2 $\times$ 10$^{-2}$  & 1.7 $\times$ 10$^{-2}$  & 2.8 $\times$ 10$^{-2}$  & 4.3 $\times$ 10$^{-2}$  & 5.2 $\times$ 10$^{-2}$ \\
 & $\Delta F_{\mathrm{p}}/F_{\mathrm{t}}$   & 1.4 $\times$ 10$^{-5}$ & 5.8 $\times$ 10$^{-4}$ & 8.3 $\times$ 10$^{-4}$ & 1.4 $\times$ 10$^{-3}$ & 2.1 $\times$ 10$^{-3}$ & 2.5 $\times$ 10$^{-3}$\\
\cline{1-8}
\\[-3ex]
\cline{1-8}
$T$ = 350 K  & $\varepsilon$ & 0.01 & 0.1 & 0.3 & 0.5 & 1.0 & 1.0+ \\
\cline{3-8}
  & $\Delta F_{\mathrm{p}}$   & 6.0 $\times$ 10$^{-22}$  & 3.8 $\times$ 10$^{-20}$  & 3.6 $\times$ 10$^{-20}$  & 6.9 $\times$ 10$^{-20}$  & 1.1 $\times$ 10$^{-19}$  & 1.0 $\times$ 10$^{-19}$\\
 &   $\Delta F_{\mathrm{p}}/F_{\mathrm{0}}$  & 2.3 $\times$ 10$^{-4}$  & 1.4 $\times$ 10$^{-2}$  & 1.4 $\times$ 10$^{-2}$  & 2.6 $\times$ 10$^{-2}$  & 4.2 $\times$ 10$^{-2}$  & 3.9 $\times$ 10$^{-2}$ \\
 & $\Delta F_{\mathrm{p}}/F_{\mathrm{t}}$   & 2.2 $\times$ 10$^{-5}$ & 1.4 $\times$ 10$^{-3}$ & 1.3 $\times$ 10$^{-3}$ & 2.5 $\times$ 10$^{-3}$ & 4.0 $\times$ 10$^{-3}$ & 3.8 $\times$ 10$^{-3}$\\
\cline{1-8}
\\[-3ex]
\cline{1-8}
$T$ = 400 K  & $\varepsilon$ & 0.01 & 0.1 & 0.3 & 0.5 & 1.0 & 1.0+ \\
\cline{3-8}
  & $\Delta F_{\mathrm{p}}$   & 4.2 $\times$ 10$^{-22}$  & 6.2 $\times$ 10$^{-20}$  & 8.6 $\times$ 10$^{-20}$  & 1.6 $\times$ 10$^{-19}$  & 1.6 $\times$ 10$^{-19}$  & 1.7 $\times$ 10$^{-19}$\\
 &   $\Delta F_{\mathrm{p}}/F_{\mathrm{0}}$  & 9.0 $\times$ 10$^{-5}$  & 1.3 $\times$ 10$^{-2}$  & 1.9 $\times$ 10$^{-2}$  & 3.5 $\times$ 10$^{-2}$  & 3.5 $\times$ 10$^{-2}$  & 3.6 $\times$ 10$^{-2}$ \\
 & $\Delta F_{\mathrm{p}}/F_{\mathrm{t}}$   & 1.4 $\times$ 10$^{-5}$ & 2.1 $\times$ 10$^{-3}$ & 2.9 $\times$ 10$^{-3}$ & 5.6 $\times$ 10$^{-3}$ & 5.6 $\times$ 10$^{-3}$ & 5.8 $\times$ 10$^{-3}$\\
 
\cmidrule[0.08em]{1-8}

\end{tabular}}

\end{table*}

\section{Discussion}
\label{Discussion}
\subsection{Conditions for Planetary Survival}
\label{Evaporation}
Many engulfed planets are not expected to survive engulfment. \citet{sok1998} proposed that planets will evaporate once they reach a location within the envelope where the sound speed exceeds the planet's escape velocity. However, as noted by \citet{oco2023}, this is not the sole criterion for dissolution---for engulfed giant planets, dissolution is instead expected to occur when the planet fills its Roche lobe \citep{rey1999, nor2006, nor2011, gui2022}. We therefore adopt this metric to determine when our simulated planets will be destroyed, and 
report the final destruction radius for each initial planet mass in Table~\ref{tab:Results}. Based on these results, we expect the planets to survive engulfment if sufficient energy is injected to dispel the envelope. However, we note that other effects not considered in this work may be important. Survival prospects are most promising when the star has evolved to the tip of the RGB or AGB, as its expansion decreases the binding energy of the stellar envelope. For instance, \citet{yar2023} found that planets as small as 10 M$_{\mathrm{Jup}}$ can eject the stellar envelope without any additional injected energy if they are engulfed at the tip of the RGB. 
Conditions for planetary survival, particularly with respect to mass and initial orbital separation, are further explored in Gaibor et al. (\textit{in prep}). 

\subsection{Can a Planet Become a Brown Dwarf?}
\label{Deuterium}
We find that a 13 M$_{\mathrm{Jup}}$ planet can accrete 0.15 $-$ 2.20 M$_{\mathrm{Jup}}$, depending on the value of \bedit{$\varepsilon$}. Allowing for super-Eddington accretion, this range extends up to 10.28 M$_{\mathrm{Jup}}$. This accretion corresponds to a final planet mass between 13.15 $-$ 23.28 M$_{\mathrm{Jup}}$, potentially sufficient to initiate deuterium burning in the planet's core and to transform the planet into a brown dwarf \citep[e.g.,][]{bur1997, bos2005, spi2011}. However, this depends on the planet's interior composition and formation history. If it formed via core accretion, it would have a large heavy-element core made of rock and ice, unsuitable for igniting fusion. As such a planet accretes mass, it would therefore become an ultra-massive planet. If the planet instead formed via gravitational instability, it may be predominantly composed of gas, and may therefore start deuterium burning (for more on giant planet formation, see reviews by e.g., \citealt{hel2014, you2025}). This suggests that it may be possible for an engulfed planet to gain sufficient mass to become a brown dwarf. We note that the feasibility of this process depends in part on the value of \bedit{$\varepsilon$}.  Additionally, if deuterium burning is ignited, this would provide an additional source of internal energy that may suppress further accretion from the stellar envelope.


\subsection{Atmospheric Characterization with JWST}
\label{JWST}
The \textit{James Webb Space Telescope} (\textit{JWST}) provides an unparalleled opportunity to study the atmospheres and dynamical histories of planets around WDs. Because of its large aperture and wavelength coverage in the mid-infrared, it is sensitive to molecules such as water, methane, carbon dioxide, and ammonia. 
It will therefore be able to test whether the abundances of WD planets are similar to Jupiter and other known gas giants, or 
if the abundance patterns are suggestive of CEE.

\begin{figure}
\centering
\includegraphics[width=\columnwidth]{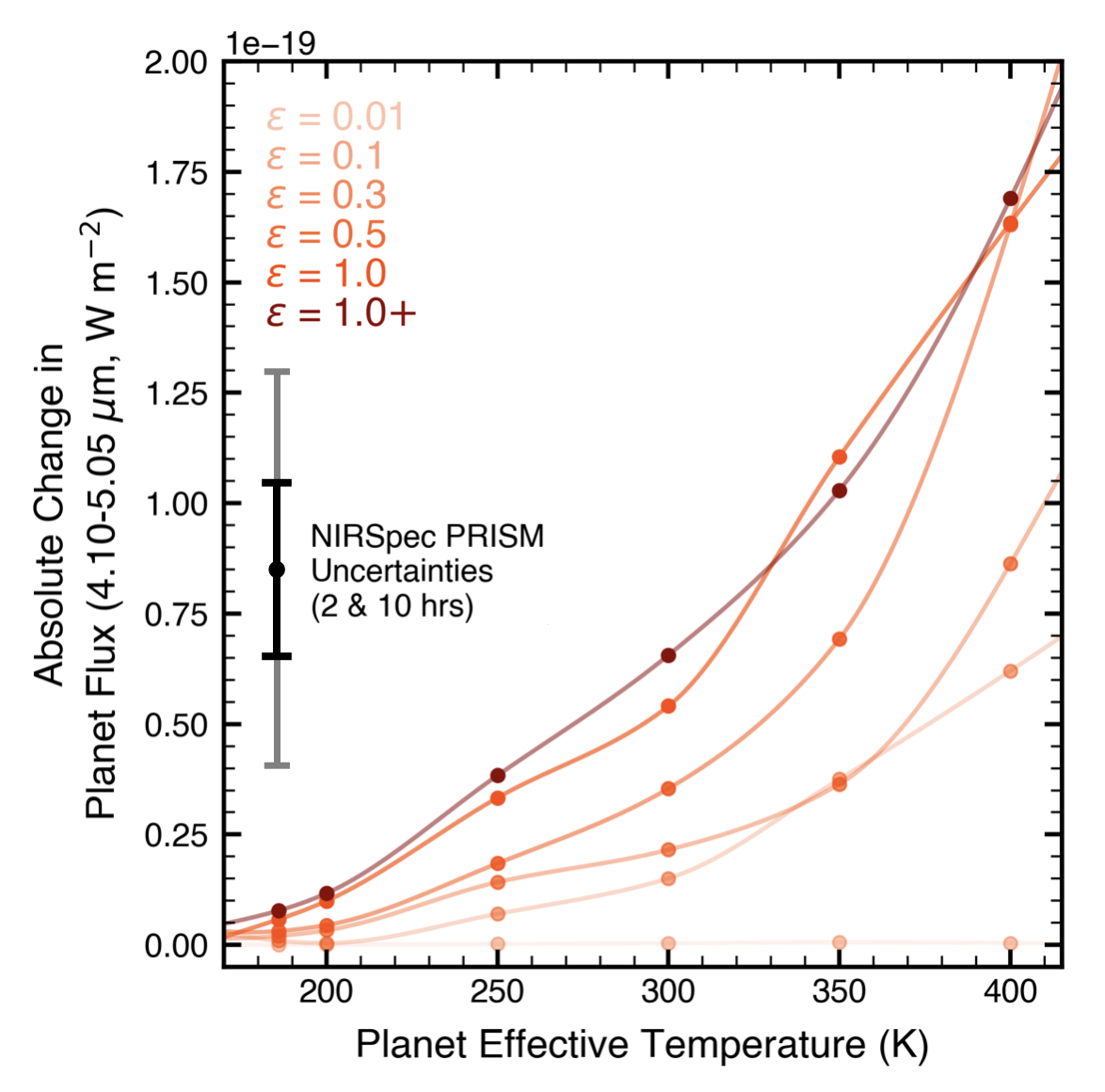}
\caption{\bedit{Absolute change} in emission of a 3 M$_{\mathrm{Jup}}$ planet \bedit{that has undergone CEE relative to a planet that has not.} 
We show the total flux in the 4.10$-$5.05 $\mu$m opacity window as a function of the planet's effective temperature\bedit{, assuming the hypothetical planet is at a distance of 24.8 pc. 
We include results for $\varepsilon\in$ [0.01, 0.1, 0.3, 0.5, 1.0], and for the scenario with an extended upper limit of 10$\times$ the Eddington limit, labeled 1.0$+$.} 
\bedit{For comparison, we} show predicted uncertainties for 2 and 10 hours of NIRSpec PRISM observations, based on Cycle 1 observations (see \S\ref{JWST}). Extended observations may be sufficiently sensitive to search for signatures of CEE on worlds hotter than WD 1856 b. \label{Temperature}}
\end{figure}

Given the small number of confirmed planets around post-main-sequence stars, there are currently few opportunities to search for such signatures. However, WD 1856 b is a prime candidate. Orbiting at just 0.0204 au from its host star, it is the closest in planet around a WD \citep{van2020}. Because WD 1856 b could not have survived its host star’s transition off the main sequence at its current position, it must have undergone some dynamical evolution either during or after the star’s transition. To better characterize this planet, $JWST$ Cycle 1 and 4 programs were accepted to observe the system with NIRSpec PRSIM \citep{prism1, prism4}. Based on the Cycle 1 results, we can measure the average flux in the 4.10$-$5.05 $\mu$m window within \bedit{$4.5\times10^{-20}$ W m$^{-2}$} with 2.0 hours of PRISM observations (Macdonald et al. \textit{in press}). With 10 hours of observations, we therefore expect to measure the system's flux with a precision of \bedit{$2.0\times10^{-20}$ W m$^{-2}$.} As shown in Figure~\ref{Composition}, this is sufficient sensitivity to begin searching for CEE accretion signatures, meaning it may be feasible to search for such signals with further observations. 
We note that other effects can impact the brightness in this window. For instance, clouds can dampen thermal emission \citep[e.g.,][]{dem2013, ste2014, ste2017}. While this may obscure any signals of CEE accretion, it also guarantees that any observed signal will not be due to cloud cover. An additional Cycle 4 program will also observe the system with MIRI MRS \citep{mrs}. MRS will probe redder wavelengths, where we expect the planet to contribute a larger fraction of the overall flux. These observations may therefore be sensitive to key features impacted by CEE, such as ammonia lines. If such signatures are detected with either NIRSpec PRISM or MIRI MRS, it will provide important insight into the planet's dynamical past. 

As more planets are discovered around WDs, we will likely find more worlds orbiting within the forbidden zone. For instance, wide-field surveys such as the Vera C. Rubin Observatory may discover hundreds of transiting worlds \citep{lun2018, cor2019}. Additional planets may also be found with microlensing \citep{bla2021, zha2024}, JWST imaging and spectroscopy \citep{lim2024, mul2024}, \textit{Gaia} astrometry \citep{per2014}, and \textit{Laser Interferometer Space Antenna} (\textit{LISA}) gravitational-wave detections \citep{tam2019}. 
If any newly discovered forbidden zone planets are hotter than WD 1856, they would be ideal targets for searching for signatures of CEE. As shown in Figure~\ref{Temperature}, we expect the signal to dramatically increase as a function of the planet's effective temperature, reducing the observing time needed to detect any such features. Searching for signs of CEE in this growing planet population will allow us to probe how stellar evolution affects planetary systems and to constrain the frequency of planetary survival in post-CEE systems. 


\subsection{Future Work} 
\label{Future work}
A limitation to our simulations is that we model the planet as a point source, and therefore do not incorporate planetary effects such as atmospheric heating or ohmic dissipation. These omitted contributions may inflate the planet's atmosphere, interrupt or terminate accretion, and drive thermal winds, potentially making it difficult for the planet to accrete and retain material. To account for these effects, future work could simultaneously model the planet's evolving atmosphere. For instance, many models have been designed to study accretion during planet formation \citep[e.g.,][]{ayl2009, mac2010, dan2013, szu2017, lam2019, sch2019_2} and to probe the atmospheres of highly irradiated hot Jupiters \citep[e.g.,][]{bur2007, arr2010, rau2012, tre2017, sai2019}. Adapting these models to a CEE scenario would provide a more realistic treatment of the relevant atmospheric physics.

A second limitation is that we assume accretion is spherically symmetric. However, given the planet's motion relative to the stellar envelope, we expect more complicated dynamics to arise. Future work could therefore refine our results by incorporating hydrodynamical simulations of the stellar envelope flow around the engulfed planet. \citet{yar2023} has previously modeled this using wind tunnel simulations that account for both gravitational and ram pressure drag. They found that when the ratio between the planet's geometrical and gravitational radii is small ($<$1), the planet gravitationally focuses gas as predicted by BHLA. Incorporating this detailed modeling of flow morphology into our accretion model would provide a more physical simulation of planet accretion. 
 
We also do not model stellar or planetary magnetic fields in our simulations. As discussed in \S\ref{Magnetic_fields}, the magnetic fields within the AGB star may be non-negligible, and the planet's magnetic field may drive important effects such as ohmic heating. Magnetic fields have previously been included in models of CEE for stellar binary systems \citep[e.g.,][]{reg1995, ohl2016, bel2019, sch2019, ond2022, gag2024}. These earlier works found that magnetic fields tend to decrease differential rotation, drive stellar winds, and launch jet-like outflows, but that they do not have a strong effect on the system's dynamics. They also found that the inspiralling companion can dramatically amplify the engulfing star's magnetic field. Extending this framework to model smaller mass companions would probe how these effects manifest for engulfed planets.

Finally, \citet{spi2012} investigated how giant planets on wide orbits may accrete material from RGB and AGB stellar winds. They found that a planet could accrete a significant fraction of its initial mass. Further work is needed to determine how accretion during CEE and accretion from winds differ, and how to differentiate between the long-term atmospheric signatures.

\section{Conclusion}
\label{Conclusion}
In this paper, we model planetary engulfment in AGB stars. Assuming Bondi-Hoyle-Littleton accretion, we calculate how much mass a planet can accrete, and predict the resulting impact on the planet's atmospheric composition. The main conclusions of the paper are summarized as follows: 

\begin{enumerate}
\item[\textbullet] Depending on the efficiency of planetary accretion, we expect planets to accrete up to 48$\%$ of their initial mass, with 3 M$_{\mathrm{Jup}}$ planets demonstrating the highest fractional increase in mass. If super-Eddington accretion is allowed, planets can nearly double their initial mass. \bedit{For low accretion efficiencies $\varepsilon\in$ [0.01, 0.1, 0.3, 0.5], we find that planets accrete 0.1$-$27$\%$ of their initial mass.}

\item[\textbullet] \bedit{Because the accreted material is enriched in hydrogen and helium, we expect it to decrease the planet's bulk metallicity, deepen opacity windows, and increase thermal emission. 
For a cool (186 K), WD 1856 b-like planet, emission in the 4.10$-$5.05 $\mu$m window increases by up to 9.0\% in the most extreme case. For low accretion efficiencies $\varepsilon\in$ [0.01, 0.1, 0.3, 0.5], we find that planet emission increases by 0.1$-$3.6\%. }

\item[\textbullet] These signatures may be observable with $JWST$. With 10 hours of NIRSpec PRISM observations, we could begin to search for signatures of CEE \bedit{on planets with favorable conditions. These 
 signatures will be most easily observable for worlds significantly hotter than WD 1856 b}.  

\end{enumerate}
Our results suggest that CEE could in favorable cases leave measurable imprints on planetary atmospheres. Searching for these CEE signatures with $JWST$ could provide one of the few constraints on the formation history of planets around WDs, and will probe the most extreme limits of planetary survival. 



\section*{Acknowledgments}
This material is based upon work supported by the National Science Foundation Graduate Research Fellowship under Grant No. 1745302. J.M. acknowledges support from the National Science Foundation Graduate Research Fellowship Program under Grant No. DGE 2137420. MSF acknowledges support for this research was provided by the Office of the Vice Chancellor for Research and Graduate Education at the University of Wisconsin--Madison with funding from the Wisconsin Alumni Research Foundation. A.V. is supported as a Sloan Research
Fellow. We thank Ruth Murray-Clay for useful discussions in the preparation of this work. 

\software{matplotlib \citep{plt}, 
          numpy \citep{np}
          }


\clearpage

\vspace{5mm}
\bibliography{refs}{}
\bibliographystyle{aasjournal}






\end{document}